\documentclass[aip,reprint,apl]{revtex4-1}
\usepackage{amsmath}
\usepackage{graphicx}
\usepackage{dcolumn}
\usepackage{booktabs}

\begin{document}
\title{Vibrational properties of graphene fluoride and graphane}
\author{H. Peelaers}
\email{hartwin.peelaers@ua.ac.be} 
\altaffiliation{Present address: Materials Department, University of California, Santa Barbara, CA 93106-5050, USA}
\affiliation{Universiteit Antwerpen, Departement Fysica, Groenenborgerlaan 171, B-2020 Antwerpen, Belgium}
\author{A. D. Hern\'andez-Nieves}
\email{alexande@cab.cnea.gov.ar} \affiliation{Universiteit Antwerpen, Departement Fysica,
Groenenborgerlaan 171, B-2020 Antwerpen, Belgium}
\affiliation{Centro Atomico Bariloche, 8400 S. C. de Bariloche, and CONICET, Argentina}
\author{O. Leenaerts}
\affiliation{Universiteit Antwerpen, Departement Fysica,
Groenenborgerlaan 171, B-2020 Antwerpen, Belgium}
\author{B. Partoens}
\affiliation{Universiteit Antwerpen, Departement Fysica,
Groenenborgerlaan 171, B-2020 Antwerpen, Belgium}
\author{F. M. Peeters}
\email{francois.peeters@ua.ac.be} \affiliation{Universiteit Antwerpen, Departement Fysica,
Groenenborgerlaan 171, B-2020 Antwerpen, Belgium}
\date{ \today }
\begin{abstract}
The vibrational properties of graphene fluoride and graphane are studied using \emph{ab initio}  calculations. We find that both {\it sp$^3$} bonded derivatives of graphene have different phonon dispersion relations and phonon density of states as expected from the different masses associated with the attached atoms fluorine and hydrogen, respectively. These differences manifest themselves in the predicted temperature behavior of the constant-volume specific heat of both compounds.
\end{abstract}
\maketitle
The chemical modification of graphene \cite{novoselov:2004} is being the focus of increasing interest.
\cite{dikin:2007,elias:2009,cheng:2010,withers:2010,nair:2010,robinson:2010,wang:2010}
Radicals such as oxygen, hydrogen, or fluorine atoms can be adsorbed on the surface of graphene.
The adsorbed radicals can attach to the graphene layer in a random way, as is the case in graphene oxide
(GO),\cite{dikin:2007} or they can form ordered patterns. The later is expected to be the case for
hydrogen\cite{sluiter:2003,sofo:2007} and fluorine adsorbates.\cite{charlier:1993,leenaerts:2010} The two-dimensional materials that
form in those cases\cite{sluiter:2003} have been named graphane \cite{elias:2009,sofo:2007}
and graphene fluoride\cite{cheng:2010} (or fluorographene\cite{nair:2010}), respectively.
\\
Raman spectroscopy provides useful information, and it is being actively used, during the hydrogenation \cite{elias:2009}
and fluorination\cite{cheng:2010,withers:2010,nair:2010,robinson:2010} process of graphene.
Both graphane and graphene fluoride are expected to be wide band gap materials.\cite{leenaerts:2010}
As the energies of the lasers used were lower than the band gap, there was no signature
of Raman activity from fully hydrogenated or fully fluorinated
regions in those experiments,\cite{elias:2009,withers:2010,nair:2010,robinson:2010} and the Raman peaks were associated with graphene.
\\
The main features in the Raman spectra of pristine graphene is the appearance of the G and 2D bands,
which appear around 1580 cm$^{-1}$ and 2680 cm$^{-1}$,
respectively.\cite{elias:2009,malard:2009} The G-band is associated with the doubly degenerate E$_{2g}$ phonon mode of
graphene, while the 2D mode (also called G') originates from a second-order process, involving two phonons near the K point
without the presence of any kind of disorder or defect.\cite{malard:2009}
On the other hand, the presence of defects in the sample (such as adatoms in partially hydrogenated or fluorinated samples)
activates additional peaks in the Raman spectra of graphene. This is the case of the D, D', and D+D' peaks that appear at 1350 cm$^{-1}$,  1620 cm$^{-1}$, and 2950 cm$^{-1}$, respectively.\cite{elias:2009, withers:2010} These Raman peaks originate from double-resonance processes at the K point in the presence of defects. As the D, D and D+D' peaks involve phonon modes from graphene, they appear at the same frequencies in partially hydrogenated\cite{elias:2009} or partially fluorinated samples.\cite{withers:2010}
\\
The D and G peaks of graphene provide valuable information on the density of defects.
These peaks tend to disappear in the limit of almost fully covered samples as was recently found in graphene
fluoride \cite{nair:2010,robinson:2010} but not in previous experiments on graphane\cite{elias:2009} or graphite
fluorides.\cite{gupta:2003,commentGF}
The first report on Raman signatures of multi-layer graphene regions that are fully covered with fluorine atoms was presented in Ref. \onlinecite{wang:2010}. By using a UV laser with an energy of 5.08 eV, two Raman active modes were detected at
1270 cm$^{-1}$ and 1343 cm$^{-1}$, which are absent for lower laser energies. An infrared active mode was also reported at 1204 cm$^{-1}$. These modes were correlated with DFT
calculations of the phonon frequencies of graphene fluoride at the $\Gamma$ point.\cite{wang:2010}
\\
In this letter, by using first-principles calculations, we investigate the phonon dispersion and the phonon density of states of graphene fluoride
and graphane. This information can be useful for the interpretation of future experiments on infrared, Raman, and
neutron-diffraction spectra of these novel two-dimensional compounds, as well as in the study of a wide variety of other physical properties such as specific heat, thermal expansion, heat conduction, and electron-phonon interaction.
\\
All reported results are obtained with the \textsc{ABINIT} code,\cite{Gonze:2009} using a plane wave basis and Troullier-Martins pseudopotentials.\cite{troullier:1991} An energy cutoff of 40 Ha was used. A 30 Bohr layer of vacuum was used to separate the graphane/fluorographene sheet from its periodical images in order to avoid unphysical interactions. The Brillouin Zone was sampled using a 30x30x1 Monkhorst-Pack grid~\cite{monkhorst:1971} for the electronic structure. Both considered systems are fully relaxed such that all forces are smaller than $5\times10^{-5}$ Hartree/Bohr and all stresses smaller than $5\times10^{-7}$ Hartree/Bohr$^3$. The dynamical properties are calculated within the density-functional perturbation theory~\cite{Baroni:2001} (DFPT), as this allows us to calculate arbitrarily phonon frequencies without requiring the use of a supercell.
\\
\begin{figure}
\includegraphics[width=0.95\linewidth]{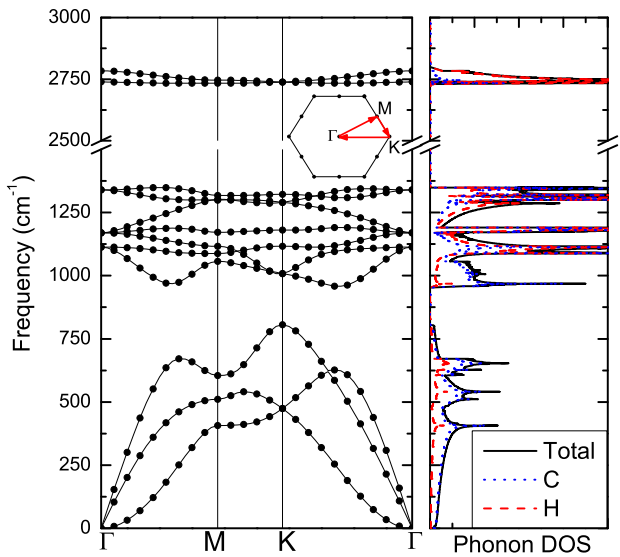}
\vspace{0cm} \caption{\label{PH_graphane} (Color online) Phonon dispersion of graphane in the chair conformation. The dots are the directly calculated frequencies and the lines are interpolated values.
The inset shows the first Brillouin zone and the wavevector path used. The right-hand side shows the phonon DOS, where the black lines indicate the total DOS, the blue dotted lines the C projected DOS, and the red dashed line the H projected DOS. }
\end{figure}
Figure~\ref{PH_graphane} shows the calculated phonon dispersion relation for the chair conformation of graphane, as this is the energetically most favourable crystal configuration.\cite{leenaerts:2010} The dots indicate the directly calculated frequencies, while the solid line is obtained by using a Fourier based interpolation of the interatomic force constants obtained on a 10x10x1 grid of {\it q}-points. As can be seen, the agreement between both calculations is very good. The absence of any imaginary frequency indicates that the chair conformation is stable. It is apparent that the phonons of graphane can be divided into low-, intermediate-, and high-frequency groups of phonons.
The right-hand side of the figure shows the corresponding density of states (DOS), obtained with the interpolated frequency values. The blue dotted line is the DOS projected on the C atoms, and the red dashed line is the projection on the H atoms.
From the projected phonon DOS, one can immediately identify the high-frequency modes as dominantly H modes, as can be expected from the C--H stretching modes. The acoustic modes are dominantly C modes while the intermediate-frequency group of phonons have significant contributions from both types of atoms.
\\
\begin{figure}
\includegraphics[width=0.95\linewidth]{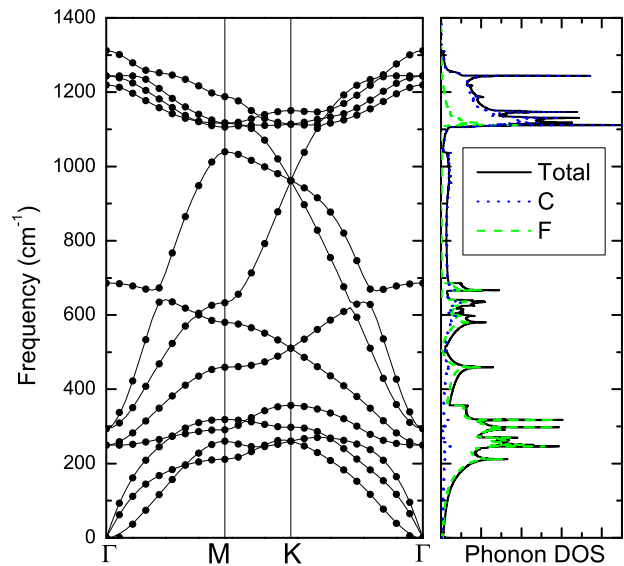}
\vspace{0cm} \caption{\label{PH_fluorographene}  (Color online) Phonon dispersion of fluorographene in the chair conformation. The dots indicate the directly calculated frequencies and the lines are the interpolated values. The right-hand side shows the phonon DOS, where the black lines indicate the total DOS, the blue dotted lines the C projected DOS, and the green dashed line the F projected DOS.}
\end{figure}
A similar approach can be applied to fluorographene, resulting in the phonon dispersion relation (left) and projected phonon DOS (right) shown in Fig.~\ref{PH_fluorographene}. In the projected DOS, the blue dotted line is the projected DOS on the C atoms, and the green dashed line is the projection on the F atoms. From the dispersion we found that also the chair conformation of fluorographene is stable.
\\
There are substantial differences in the phonon spectra and DOS of graphane and fluorographene. The latter does not show clearly separated groups of phonons. The high-frequency modes are also lower in frequency in comparison with similar modes in graphane because F is heavier than H. In this case, the dominant contribution to the acoustic modes comes from the F atoms while the high-frequency modes around 1200 cm$^{-1}$ have a clear C character. This behavior is the opposite from the one in graphane because the mass of C atoms is in between the mass of F and H atoms.
\begin{table}[htbp]
\centering
\caption{List of symmetries and phonon frequencies at different {\it q}-points for graphane and fluorographene. R and I stand for Raman and infrared active modes, respectively.\\}
\begin{tabular}{c|ccc|ccc}
\addlinespace
\toprule
\hline
\hline
Symmetry& \multicolumn{3}{c|}{Graphane} &  \multicolumn{3}{c}{Fluorographene} \\
\midrule
(activity)& $\Gamma$ & M     & K     & $\Gamma$ & M     & K \\
\hline
A$_{2u}$\;(R)& 0     & 407   & 474   & 0     & 211   & 260 \\
E$_u$\;(I) & 0     & 511   & 474   & 0     & 260   & 260 \\
E$_u$\;(I) & 0     & 605   & 806   & 0     & 291   & 298 \\
E$_g$\;(R) & 1114  & 1057  & 1008  & 250   & 318   & 356 \\
E$_g$\;(R) & 1114  & 1088  & 1008  & 250   & 459   & 510 \\
E$_u$\;(I) & 1170  & 1116  & 1117  & 294   & 580   & 510 \\
E$_u$\;(I) & 1170  & 1172  & 1180  & 294   & 633   & 961 \\
A$_{1g}$\;(R) & 1170  & 1299  & 1290  & 686   & 1039  & 962 \\
E$_g$\;(R) & 1339  & 1302  & 1290  & 1244  & 1106  & 962 \\
E$_g$\;(R) & 1339  & 1317  & 1322  & 1244  & 1116  & 1114 \\
A$_{2u}$\;(I) & 2739  & 2733  & 2739  & 1219  & 1117  & 1114 \\
A$_{1g}$\;(R) & 2783  & 2746  & 2739  & 1312  & 1188  & 1150 \\
\hline
\hline
\bottomrule
\end{tabular}
\label{tab:freq}
\end{table}
\\
The 12 phonon modes of the chair conformation of graphane and graphene fluoride, which has the \emph{P-3m1} symmetry (space group 164), belong to 4 irreducible representations. The frequency of the different phonon modes at three high-symmetry points of the Brillouin Zone ($\Gamma$, $K$, and $M$) are summarized in Table \ref{tab:freq}. E$_g$ and A$_{1g}$ are Raman active modes while E$_u$ and A$_{2u}$ are infrared active modes.
Our theoretical values of the A$_{1g}$ and the two-fold degenerate E$_g$ modes of
graphene fluoride at the $\Gamma$ point, 1244 cm$^{-1}$ and 1312 cm$^{-1}$ are in good agreement
with the two Raman active modes found experimentally in Ref.~\onlinecite{wang:2010} at 1270 cm$^{-1}$ and 1345 cm$^{-1}$, respectively.
The same agreement is observed between the infrared active mode A$_{2u}$ at 1219 cm$^{-1}$ and the report of an
infrared active mode at 1204 cm$^{-1}$ in Ref.~\onlinecite{wang:2010}.
\\
\begin{figure}
\includegraphics[width=0.7\linewidth]{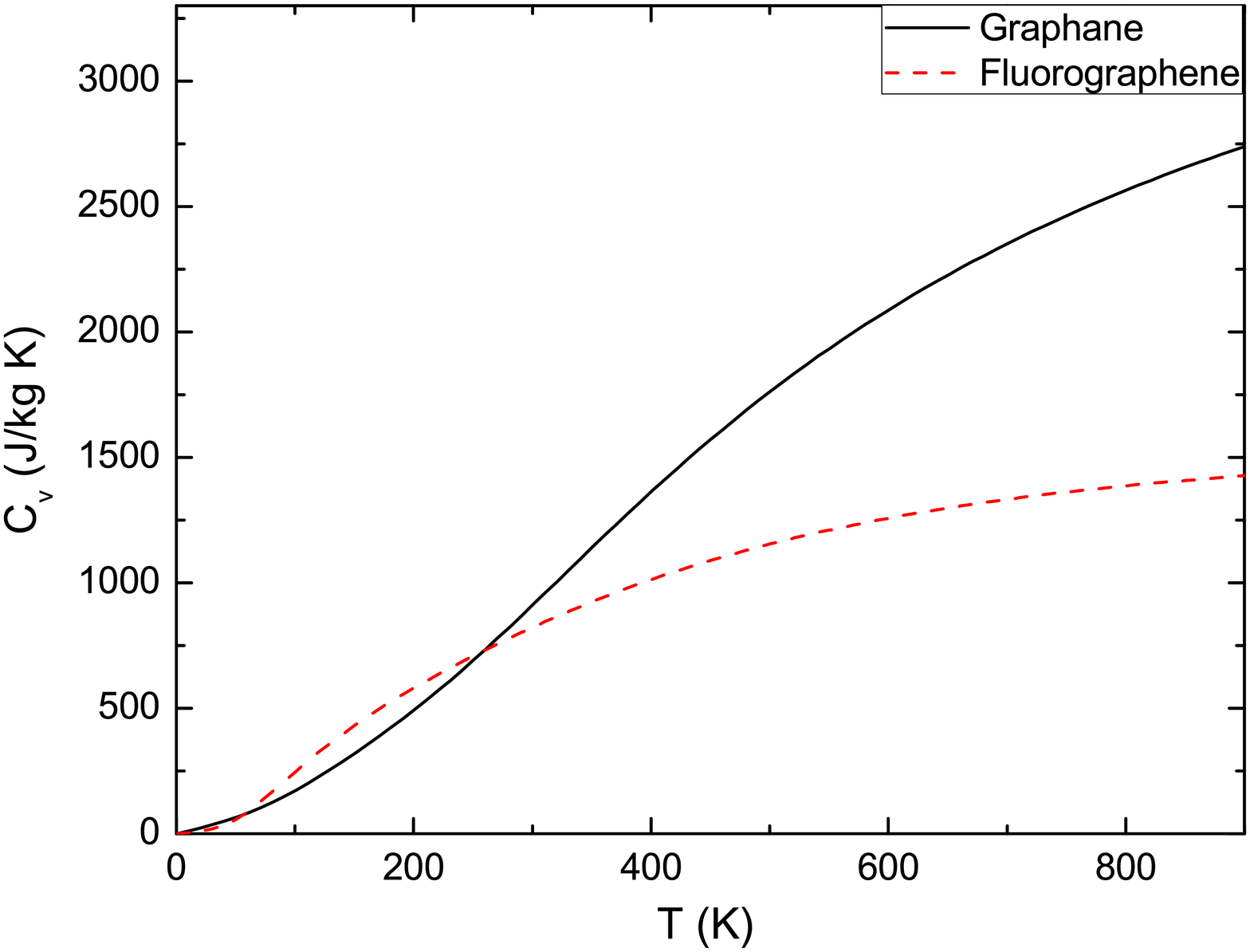}
\vspace{0cm} \caption{\label{C_v}  (Color online) The constant-volume specific heat of graphane (black solid line) and fluorographene (red dashed line). At high temperature, the specific heat of graphane is much larger than the one of fluorographene.}
\end{figure}
Using the phonon DOS we can also calculate the constant-volume specific heat of both materials. This is achieved using the expressions described in Ref.~\onlinecite{Lee:1995}. One should keep in mind that we work within the harmonic approximation, so the differences with experimental values will become larger for higher temperatures. The hydrogenation of graphene was found to be reversible at 700 K in Ref. \onlinecite{elias:2009} while the fluorination of graphene could be reverted at 870 K in Ref. \onlinecite{cheng:2010}. The calculated values of C$_v$ are shown in
Fig.~\ref{C_v} where we only report the behavior in the region of stability of each compound.
The specific heat of graphane is larger than the one for fluorographene, as can be expected considering that the classical limit of the specific heat is given by the Dulong-Petit law and is equal to $3R/M$, where $R$ is the ideal gas constant and $M$ the molar mass. Because the molar mass of fluorographene is larger than that of graphane, the value of the specific heat will be smaller at high temperature.
To our knowledge there are no experimental measurements yet to compare with. The specific heat of graphene at high-temperature [see Fig. 24 of Ref. \onlinecite{Mounet:2005}] is in between the specific heat of graphene fluoride and graphane as expected from the Dulong-Petit law.
\\
In summary, the experimental determination of the phonon dispersion relations of graphane and fluorographene can be very useful
in the characterization of these materials. It is worth noting, that both graphene derivatives are wide band gap materials.
This allows to clearly discriminate, for example, the Raman activity originated in fully-fluorinated and fully-hydrogenated regions during the synthesis of both compounds by using a laser with appropriate energies.

\begin{acknowledgments}
This work was supported by the Flemish Science Foundation (FWO-Vl), the Belgian Science Policy (IAP), and the collaborative project FWO-MINCyT (FW/08/01). A. D. H. is also supported by ANPCyT (under Grant No. PICT2008-2236).\\
\end{acknowledgments}

\end{document}